# From Robots to Books: An Introduction to Smart Applications of AI in Education (AIEd)

Shubham Ojha [1]   Siddharth Mohapatra[1]   Aditya Narendra[1]   Ipsit Misra[1]

[1] Center of Excellence of Artificial Intelligence, Odisha University of Technology and Research Bhubaneswar

ABSTRACT:

The world around us has undergone a radical transformation due to rapid technological advancement in recent decades. The industry of the future generation is evolving, and artificial intelligence is the next change in the making popularly known as Industry 4.0. Indeed, experts predict that artificial intelligence (AI) will be the main force behind the following significant virtual shift in the way we stay, converse, study, live, communicate and conduct business. All facets of our social connection are being transformed by this growing technology. One of the newest areas of educational technology is Artificial Intelligence in the field of Education (AIEd). This study emphasis the different applications of Artificial Intelligence in education from both an industrial and academic standpoint. It highlights the most recent applications of AIEd, with some of its main areas being the reduction of instructors' burden and students' contextualized learning novel transformative evaluations, and advancements in sophisticated tutoring systems. It analyses the AIEd's ethical component and the influence of this transition on people, particularly students and instructors as well. Finally, the article touches on AIEd's potential future research and practices. The goal of this study is to introduce the present-day applications to its intended audience.

I. Introduction:

The emergence of artificial intelligence(AI) has created a tremendous technical change in recent years.AI as defined by Marvin Minsky, artificial intelligence(AI) is the science of making machines do things that would require intelligence if done by men. This topic began as a research area of computer science engineering, but due to its significant absorption of ideas from neurology, cognitive science, philosophy and other disciplines, it has become extremely interdisciplinary, making it difficult even for experts to find an agreeable definition of artificial intelligence. It is a system that has capabilities(such as language or perception) and intelligent conduct that were originally thought to be exclusive to mankind and carry out a certain task. In simpler terms, artificial intelligence (AI) is a discipline of computer science dealing with the emulation of human intelligence by behaving intelligently. This formidable technology has brought about a shift in the way we live in the world.

By incorporating AI-based solutions, sectors including manufacturing, healthcare, etc, are undergoing a sea of change in their operational methodologies. Around the world, the education sector exhibits a similar pattern. Given the largely advantageous digital changes that AI brings into the system, AI has undoubtedly created challenges to traditional ways of education. For the past 30 years, researchers have been studying integration of artificial intelligence in education(AIEd).AIEd has achieved significant success in strengthening connections between teachers and student where the connections were lacking or needed improvement. With the use of AI effective teaching techniques, evaluation systems, and feedback mechanisms can also be introduced. Additionally, weaknesses in the existing systems can be identified, and variety of student responses like boredom and concentration can be captured to make learning a interactive environment.

This essay provides a survey of the most recent advancements in artificial intelligence in education. It starts out by going over numerous fields of education and learning that have made us of AI, then shifts to the areas on which we see the industry concentrating, and it ends with a remark on further fields of development with AI in Education providing a succinct overview of the domain.

II. Methodology:

This multi-phase study conducts a comprehensive analysis of peer-reviewed articles on AI in education. To discover eligible publications for complete analysis, a multi-phase search and selection process was used.

A. Archival Databases

Using the proliferation of online journals and freely accessible resources, even with well-defined criteria, it's nearly difficult to do a thorough search. This study was methodically planned to concentrate on research articles gathered within one of the most commonly used web-based databases, Google Scholar (SCI/SSCI). The database used as the source was chosen as It compiles publications from the Social Science Citation Index (SSCI) and the Science Citation Index (SSCI). Besides that, Possibly not including more recent journals in one of the Science database, further searches were carried out to find newly published papers on newly published papers. Furthermore, given the interdisciplinary nature of AIED, relevant research is frequently published in more conferences on AI and learning science such as the Conference and Workshop on International Conference on Learning Representations (ICLR), International Conference on Machine Learning (ICML), International Conference of the Learning Sciences, and Neural Information Processing Systems (NEURIPS) (ICLS).

B. Criteria for Searches and Selection

The source database was searched several times and carried out employing a variety of essential crucial phrase assemblages and search tactics, such as "AI," "artificial intelligence," and "education.". Two non-English articles and 27 duplicates were eliminated during the first screening of the 307 items that were produced after 18 rounds of Web of Science searches. Furthermore, looks for supporting information to study screening were undertaken on the websites. To achieve the goals of the research, first step were taken to sort the papers according to their year of publication. Second step was taken to find subdomains that were mainly covered in the papers. It was also discovered that the most recent AIEd research can be separated into four major subdomains, as illustrated in fig 2. which were chosen as research and study subjects for the paper .In order to meet the research objectives, a set of inclusion and exclusion criteria were developed and applied. For subsequent analysis, only English language refereed journal papers providing empirical, evidence-based investigations were chosen. The methodology for the paper analysis and selection process is given below in Fig 1.

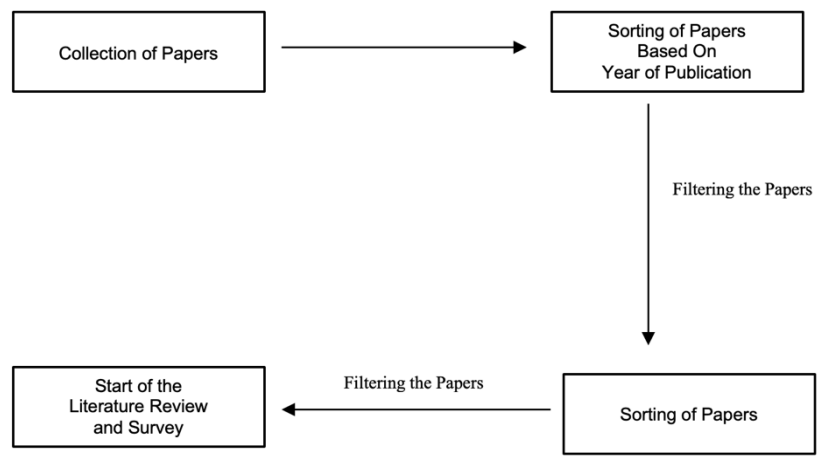

Fig 1. The Methodology of Paper Analysis and Selection Process

TABLE I. SELCTED SUB DOMAINS OF APPLICATIONS OF AI FOR EDUCATION CHOSEN FOR THIS STUDY

| Sub Domains of Work | Purpose of Working |
| --- | --- |
| Teacher's / Instructor's Workload Decrement | To lessen the amount of work teachers have to do without affecting student learning |
| Adaptive Curriculum | On the basis of students' contexts and learning histories, offer them personalised and/or customised learning experiences. |
| Intelligent Tutoring Systems | Construct educational scenarios that engage students, offer individualised feedback, and improve their comprehension of particular subjects. |
| Smart Assessments | Increasing understanding of students which covers not just what people are cognizant about, but also how they learn and which pedagogies are most effective for them, thanks to the use of adaptive assessment. |

III. AI for Education Applications

Virtually endless options in education fields are made possible by AI technology. The following categories of learning technology were examined by the twenty studies as part of a range of educational AI applications.

    A. Teacher's/ Instructor Workload Decrement( [1] [2] [3] [4] [5] )
    B. Adaptive Curriculum ( [6] [7] [8] [9])
    C. Intelligent Tutoring Systems([10] [11] [12] [13] )
    D. Smart Assessments([14] [15] [16] [17])

A. Teacher's / Instructor's Workload Decrement

With good reason, recent AIED research has prioritised teachers over other educational institution stakeholders. No matter the type of learning environment—face-to-face or online—teachers are at the centre of it. Utilizing participatory design methodologies, new AIED technologies are developed with teachers, parents, and students in mind. In order to empower teachers and provide them the freedom to concentrate on teaching rather than other responsibilities, educators have long faced the difficulty of reducing teacher burden.

Teachers will need to make adjustments more frequently as a result of the widespread emphasis on online learning and the creation of new tools to support it. It is essential for teachers to upgrade and retrain themselves in order to adapt to this generation, particularly the new skills they must pick up in order to fully profit from AIED. In order to understand, evaluate, and adapt to new educational technological tools as they become available, people must first become digitally literate. They might or might not use these tools, but it's important to know what they offer and whether or not they help instructors to reduce their burden. Zoom video calling was frequently used to deliver classes remotely throughout the outbreak. In addition to setting up courses on Zoom, teachers also need to understand how to use the breakout rooms for group work and the whiteboard for free-hand writing. In order to evaluate the data supplied by these ed-tech tools and to decide what kind of data and analytics tool they will need to better understand students, teachers will also need to develop their analytical skills. As a result, teachers will be able to buy only the ed-tech products they require, which will lighten their workload. Finally, in order to incorporate new tools into their daily practises, educators will need to master new collaborative, group, and management abilities. They will be responsible for efficiently managing these extra resources.

In recent studies, Talk Moves[1], an application designed to give teachers feedback and incite fruitful dialogues in their maths courses, was introduced. This application employs models such as BERT and LSTM paired with BOW with Glove to provide individualised feedback based on recordings. Another study introduced a Neural Geometric Solver[2] that is capable of automatically solving geometric problems trained using a dataset named GeoQA; the study also proposed its usage of jigsaw location prediction, geometric elements prediction, and knowledge points prediction using neural nets and LSTM as a decoder-encoder pair. The vast majority of work has also been completed for handwriting text validation and document conversion.

A study by Singh and Karayev[3] developed a model for automatic handwriting text identification from images and translation into text, all done in sequence using neural networks and transformers which may be used for easy document generation, including notes, and easy assignment verification. PREREQ[4] is a solution for Pre-Requisite Annotation and identifying proper background information when growing online education technology. It is a supervised learning strategy that efficiently learns the prerequisites for online educational courses from videos and other resources. OCR is a well-studied topic, but latex-based OCR is a new concept described in a recent study[5] that uses neural nets to directly output text from a latex-based document, which can help educators decode research papers and keynotes to text while saving time and effort. A brief overview of these works are given below in TABLE II.

TABLE II. AN OVERVIEW OF WORKS COVERED UNDER INSTRUCTOR'S WORKLOAD DECREMENT

| Name of the Work | Approach | Dataset | Use Cases |
| --- | --- | --- | --- |
| Talk Moves Application[1] | Uses SOA-NLP techniques to provide feedback to teachers | Self-Collected Datasets with 175757 sentences of each type | Can be used for providing new method for feedback collection |
| GeoQA[2] | LSTM and ResNet-101 as encoder-decoder for solving geometric questions | GeoQA, a dataset with 4998 geometric issues, answers questions about geometry. | Can be Used for doubt clearing and reduce teacher's workload |
| Full Page Handwriting Recognition[3] | Neural Network based Handwritten Text Recognition model | IAM dataset- Benchmark for handwriting recognition | Can be used for conversion of handwritten textual information into digital format or vice-versa |
| PREREQ[4] | Supervised Learning Method for finding concept wise pre-requisite relations | University Course Dataset(Benchmark Dataset) and a new self-created NPTEL MOOC Dataset | Can be used for teacher's to easily find pre requisites for each concept and create class discourse accordingly |
| What you Get is What you see [5] | Deep Learning System for OCR for latex based documents | IM2LATEX-100K dataset with rendered mathematical expressions from published papers | Used by teachers to easily create resources from latex based documents. |

B. Adaptive Curriculum

Based on their social background, level of financial security, mental stability, and prior knowledge of the subject, each learner has a unique learning environment. When instruction is tailored to these changing circumstances, it is most successful. The use of AIED can help identify a learner's specific learning gaps, suggest appropriate reading material based on those gaps, and offer detailed solutions to challenging issues. These tailored learning plans not only allow for simple development of students into higher level ideas, but also ensure that the fundamentals of each concept are carefully read and taken care of. Aside from that, tailor-made educational curriculum enables institutions to build more individualised programmes with greater student participation with the teacher, increasing the entire programme experience.

For instance, researchers created the open-source platform iTalk2Learn[6] to help students in grades 5 through 11 learn mathematics more effectively. This tutor spoke with the students, identifying when a student was having trouble with fractions and conducting the appropriate interviews.. Another latest study by Alkadi and Inkpen[7] established a model that classifies student reading materials based on their readability using multiple machine learning algorithms, which can help educators aid readers with different reading levels in selecting resources that fit their readability level. A different introduced making changes to general taxonomy books PDF2PreReq[8] introduced dynamic textbook algorithms that link necessary concepts with each topic and chapter, saving time and producing a better experience.

Student frustration can frequently lead to delays in concept formation and understanding. In order to address contemporary curriculum and interview settings, Betty's Brain[9], an interviewing agent, captures these irritation lapses. A brief overview of these works are given in below TABLE III.

TABLE III. AN OVERVIEW OF WORKS COVERED UNDER ADAPTIVE CURRICULUM

| Name of the Work | Approach | Dataset | Use Cases |
|---|---|---|---|
| iTalk2Learn[6] | Smart Curriculum Based Maths Tutoring | Not Available | For Adaptive Curriculum setting based on student' need and expertise to improve interest and knowledge |
| Classifying Documents based on Multiple Readability Levels[7] | Deep Learning based system for Classification of articles | Newsela Dataset | For personalized article recommendation based on expertise |
| PDF2PreReq [8] | End to End pipeline for generating dependency graphs for existing curriculum | Not Available | For finding dependency on underlying concepts and smart curriculum setting |
| Affect-Targeted Interviews for Understanding Student Frustration[9] | Analysis of various student frustration by use of virtual agent during studying | Self-Collected Dataset through student interviews | For Understanding perspective of student during study process and set curriculum accordingly |

C. Intelligent Tutoring Technologies

A computer programme known as an intelligent tutoring system makes an effort to communicate with a human teacher in order to give students individualised instruction. The concept of ITS in AIED has existed for some time. High standards have always been set for ITS's capacity to support education. IT has been noticed a considerable gap between what ITS were promised to accomplish and what they've actually been instrumental in providing throughout the years.

The majority of ITS in recent years have been subject- and subject-focused, like ASSISTments[10]. Each intelligent tutoring system focuses on a certain subject, however results has shown that they are effective at giving students relevant information, interacting with students, and improving students' academic achievement.

An Intelligent Question Answering System[11] was proposed in a different study. It combines intelligent systems for answering questions based on knowledge graphs, incorporates big data technology, and uses them to quickly and accurately answer questions from high school students while connecting the knowledge points that are relevant to the questions. It also analyses students' questioning behaviour and anticipates student learning behaviour to offer information on the impact of instruction.

In a work that combined BASEBERT models with domain adaptive pre training, a generic language model[12] for question-answering was proposed that surpassed simultaneously with the most recent state-of-the-art model. A neural model[13] with visual attention was developed in a different study, and it can be trained to learn how to mark up a mathematical formula in LaTeX from its image. This paradigm can be used to teach computers various coding methods. A brief overview of these works are given in below TABLE IV.

TABLE IV. AN OVERVIEW OF WORKS COVERED UNDER INTELLIGENT TUTORING SYSTEMS

| Name of the Work | Approach | Dataset | Use Cases |
|---|---|---|---|
| ASSITments[10] | Intelligent Ecosystem of teachers and students to provide assessments and assistance together | Self-made dataset of individual modules with Q/A, videos etc. | Provides Immediate feedback to student's on their work and analyses data to students. |
| The creation and study of an intelligent question-answering system. [11] | Knowledge graph-based big data system for answering questions | Not Available | For feedback based Q/A ensuring better student studying experience |
| MathBERT [12] | BERT based language model for maths education | Dataset of large mathematical corpus of different study levels including mathvocab | Used for automatic scoring systems based on maths assesments at various online platforms |
| Teaching Machines to Code [13] | Neural Transducer model for maths formula to latex generation | I2L-140k, Im2latex-90k | Can be used for setting student supportive personalized study plans. |

D. Smart Assessments

Any assessment of a student's work or performance in a classroom (or any type of judgement or evaluation) is referred to as assessment . Assessments, together with curriculum, learning, and teaching, are described as one of the three pillars of schooling by Hill and Barber [28]. Today's generation exams are designed to evaluate students' knowledge, comprehension, and skills. The best tests assessment platforms would consider the whole spectrum of student skills and offer useful data on learning outcomes. However, each learner and their learning process are unique. One issue that is brought up in relation to more general notions of educational evaluation is how standardised assessments may be utilised to evaluate each student, who has unique capacities, passions, and expertise.

Researchers at UCL Knowledge Lab created an intelligent assessment tool called AI Assess as an illustration [14]. Three models—the knowledge model, the analytics model, and the student model—were used to assess students' math and science learning. Each topic's information was stored in the knowledge component, student progress was monitored using the student model, and interactions between students were examined using the analytics component. Similar to this, Samarakou et al. developed an AI assessment tool[15] that evaluates students qualitatively in order to relieve teachers of the burden of spending hours analysing each activity. Applying ML methods such as RL, semantic analysis, voice recognition, and NLP can raise the calibre of assessments conducted with these technologies.

Another study published by Huang and Li [16] introduced BERTEdu, a Chinese education-based pre-trained language model that discovers similar tasks based on input by capturing semantics, diverse texts, and formulas from the exercise. A recent study uses Knowledge Tracing to propose a broad framework[17] that takes into account student engagement, the amount of difficulty of instructional activities, and natural language processing embeddings of each concept's text and predicts future performance of a student using neural nets. A brief overview of these works are given in below TABLE V.

TABLE V. AN OVERVIEW OF WORKS COVERED UNDER SMART ASSESMENTS

| Name of the Work | Approach | Dataset | Use Cases |
|---|---|---|---|
| AIAssess[14] | AI based assessment system which provides adaptive tasks based on student's progress | Not Available | For Adaptive Curriculum setting based on student' need and expertise to improve interest and knowledge |
| Implementation of AI Assessments in Engg. Lab Education[15] | Cognitive theory based assessments with feedback ensuring qualitative evaluation | Experimentative MATLAB course based dataset | For providing improvised evaluation and testing in diverse coursework. |
| An Empirical Study of Finding Similar Exercises [16] | BERT based model to find similar exercises in various fields of education | Self-Created Dataset involving education based corpus | For finding similar exercises and enabling easy assessments setting |
| Deep Knowledge Tracing using Temporal Convolutional Networks[17] | Language and neural network based model for tracing knowledge of students in assessments | Algebra 2007-2008 Dataset | For planning and setting assessments based on students' knowledge level and improve setting process. |

## IV. DISCUSSION

The field of AI for education must continue to advance in order to benefit not only education (increasing inclusivity, helping students grasp complex ideas, and increasing accessibility), but also the development of reasoning-capable AI systems. The development of AI systems that support human capacities should be considered in addition to the development of autonomous AI systems. To construct robust AI systems in the field of education, study must be done in the following areas: (a)Similar to the computer vision field, we need better benchmarking datasets such as ImageNet[39], COCO[40]. Curation of consensus benchmark dataset will provide researchers with a consistent baseline for their model, allowing them to enhance their model further. (b) We must create AI systems whose thought processes are transparent, i.e., the systems must be able to articulate their thought processes and have those processes be understandable by humans. (c) Many STEM fields require the capacity to digest knowledge from various modalities, including texts and visuals. As a result, much development in multimodal DL is required. (d)In the past, symbolic approaches to stem fields produced positive outcomes. Their key strength is their ability to reason in an interpretable manner. However, they are not scalable. So much research is needed to figure out how to combine ML and symbolic methods to AI, i.e. near symbolic AI that can ensure better AI for Education systems.

In order to examine AIED over longer time periods and at the institutional, regional, and national levels, research must be broadened. Utilizing cutting-edge technology like text mining, learning analytics, and data visualisations is also required to progress AIED research. Emerging educational research approaches are ideal for investigations on revolutionary technology like AIED, in particular educational design research (EDR), are strongly advised because they enable educators to integrate their research inquiries as part of the technology development and implementation cycle in real-world settings. When educators help with the competition, development, or evaluation of AI technology for educational objectives, EDR can be particularly beneficial. Among the range of AIEDs accessible, some have drawn more attention from researchers than others, For example, the review discovered that while very few academic publications mentioned the use of chatbots or ML in educational field, the majority of studies concentrated on intelligent tutors or personalised learning environments. Therefore, more AIED technologies should be covered in future research, especially ones that have gotten less attention over time.

## V. CONCLUSION

The motive of the paper was to highlight key technologies in AI that will create a great impact in the educational space by filling the gaps that persist in today's education space. The main highlights of the paper are to make readers cognizant of various problems that exist in today's education system, what technologies exist in today's AI that can fill those gaps, and in what manner today's AI technologies lack the capability to build next-generation AI educational tools. An in-depth analysis of the literature was used as part of a qualitative research study. AI in education has great potential and has the ability to completely revolutionize the educational space but its full potential has not been fully realized. Various stakeholders like technocrats, politicians, teachers, students, and others should collaborate on ideas to harness AI's full potential in the educational space. As AI in education can have both positive and negative side effects so the various stakeholders should correct calculations about its trade-off before deploying AI technologies in the real world. The main aim of AI is not to replace various stakeholders but to empower them. With proper policy making and careful implementation AIEd is going to prove an game changer.

## VI References

[1] Suresh, Abhijit & Jacobs, Jennifer & Lai, Vivian & Tan, Chenhao & Ward, Wayne & Martin, James & Sumner, Tamara. Using Transformers to Provide Teachers with Personalized Feedback on their Classroom Discourse: The TalkMoves Application.(2021). https://doi.org/10.48550/arXiv.2105.07949

[2] Jiaqi Chen, Jianheng Tang, Jinghui Qin, Xiaodan Liang, Lingbo Liu, Eric P. Xing, Liang Lin.GeoQA: A Geometric Question Answering Benchmark towards Multimodal Numerical Reasoning. (2022) .https://doi.org/10.48550/arXiv.2105.14517

[3] Sumeet S. Singh, Sergey Karayev,Full Page Handwriting Recognition via Image to Sequence Extraction, (2022), https://doi.org/10.48550/arXiv.2103.06450

[4] S. Roy, M. Madhyastha, S. Lawrence, and V. Rajan, "Inferring Concept Prerequisite Relations from Online Educational Resources", *AAAI*, vol. 33, no. 01, pp. 9589-9594, (2019).

[5] Deng, Y., Kanervisto, A., & Rush, A.M.What You Get Is What You See: A Visual Markup Decompiler. (2016). *ArXiv, abs/1609.04938*.

[6] Grawemeyer, B., Gutierrez-Santos, S., Holmes, W., Mavrikis, M., Rummel, N., Mazziotti, C., Janning, R.: Talk, tutor, explore, learn: intelligent tutoring and exploration for robust learning, p. 2015. AIED, Madrid (2015)

[7] Alkadi.,InkPen.,Classifying Documents to Multiple Readability levels., In AAAI2021 Spring Symposium on Artificial Intelligence for K-12 Education,(2021)

[8] Rushil Thareja, Venktesh V and Mukesh Mohania. Pdf2PreReq: Automatic Extraction of Concept Dependency Graphs fromAcademic Textbooks. In AAAI2022 Artificial Intelligence for Education,(2022)

[9] Ryan S. Baker, Nidhi Nasiar, Jaclyn L. Ocumpaugh, Stephen Hutt, Juliana M. A. L. Andres, Stefan Slater, Matthew Schofield, Allison Moore, Luc Paquette, Anabil Munshi, and Gautam Biswas.Affect-Targeted Interviews for Understanding Student Frustration. In Artificial Intelligence in Education: 22nd International Conference(2021)

[10] Heffernan, N.T., Heffernan, C.L. The ASSISTments Ecosystem: Building a Platform that Brings Scientists and Teachers Together for Minimally Invasive Research on Human Learning and Teaching. Int J Artif Intell Educ 24, 470–497 (2014) https://doi.org/10.1007/s40593-014-0024-x

[11] Zhijun Yang, Yang Wang, Jianhou Gan, Hang Li, and Ning Lei. 2021. Design and Research of Intelligent Question- Answering(Q&A) System Based on High School Course Knowledge Graph. Mob. Netw. Appl. 26, 5 (Oct 2021), 1884–1890. https://doi.org/10.1007/s11036-020-01726-w

[12] Jia Tracy Shen, Michiharu Yamashita, Ethan Prihar, Neil Heffernan, Xintao Wu, Ben Graff, Dongwon Lee.  MathBERT: A Pre-trained Language Model for General NLP Tasks in Mathematics Education,In NeurIPS 2021 MATHAI4ED Workshop.(2021). https://doi.org/10.48550/arXiv.2106.07340

[13] Singh, Sumeet.Teaching Machines to Code: Neural Markup Generation with Visual Attention. (2018). https://doi.org/10.48550/arXiv.1802.05415

[14] Luckin, Rosemary. (2017). Towards artificial intelligence-based assessment systems. Nature Human Behaviour. 1. 0028.10.1038/s41562-016-0028https://doi.org/10.48550/arXiv.1310.3174

[15] Samarakou, M., Fylladitakis, E., Prentakis, P., Athineos, S.: Implementation of artificial intelligence assessment in engineer-inglaboratory education. https://files.eric.ed.gov/fulltext/ED557 263.pdf (2014).

[16] Tongwen Huang, Xihua Li (2021) An Empirical Study of Finding Similar Exercises. https://doi.org/10.48550/arXiv.2111.08322

[17] Ait Khayi, N. *Deep Knowledge Tracing using Temporal Convolutional Networks*. *Proceedings of the Workshop ArtificialIntelligence for Education (IJCAI 2021))*, (2021). Retrieved from https://par.nsf.gov/biblio/10290861.